\documentclass[conference]{IEEEtran}
\IEEEoverridecommandlockouts

\usepackage{cite}
\usepackage{amsmath,amssymb,amsfonts}
\usepackage{algorithm}
\usepackage{algpseudocode}
\usepackage{graphicx}
\usepackage{grffile} 
\usepackage{textcomp}
\usepackage[table, svgnames, dvipsnames]{xcolor}
\usepackage{verbatim}
\usepackage{braket}
\usepackage{subcaption}
\usepackage{placeins}
\usepackage{multirow}
\usepackage{booktabs}
\usepackage{makecell}
\usepackage{tikz}
\usetikzlibrary{arrows.meta,positioning,fit,calc,quotes}
\usepackage{hyperref}
\definecolor{darkgreen}{RGB}{0,100,0}
\definecolor{tableheaderdark}{RGB}{17,24,39}
\definecolor{tableheaderibm}{RGB}{204,85,0}
\definecolor{tableheaderionq}{RGB}{75,85,99}
\definecolor{metricrowa}{RGB}{219,234,254}
\definecolor{metricrowb}{RGB}{224,231,255}
\definecolor{winnergreen}{RGB}{220,252,231}

\definecolor{tableheaderdark}{HTML}{243B53}
\definecolor{tableheaderibm}{HTML}{2F6690}
\definecolor{tableheaderionq}{HTML}{C96A3D}
\definecolor{tableheaderatom}{HTML}{5E8C61}
\definecolor{metricrowa}{HTML}{EEF4F8}
\definecolor{metricrowb}{HTML}{F8FBFD}
\definecolor{sectionrow}{HTML}{D9E6F2}
\definecolor{winnergreen}{HTML}{D7F0D1}

\newcommand{\metricwinner}[1]{\cellcolor{winnergreen}\textbf{#1}}

\def\BibTeX{{\rm B\kern-.05em{\sc i\kern-.025em b}\kern-.08em
    T\kern-.1667em\lower.7ex\hbox{E}\kern-.125emX}}

\begin{document}

\title{ InterQ: Communication-Aware Scheduling Across Modular QPUs with Classical and Quantum Links
}
\author{
\IEEEauthorblockN{
Vinooth Kulkarni\IEEEauthorrefmark{1},
Jaehyun Lee\IEEEauthorrefmark{1},
Lauren Li\IEEEauthorrefmark{2},
Aaron Orenstein\IEEEauthorrefmark{1}
}
\IEEEauthorblockN{
Xinpeng Li\IEEEauthorrefmark{1},
Shuai Xu\IEEEauthorrefmark{1},
Daniel Blankenberg\IEEEauthorrefmark{3},
and Vipin Chaudhary\IEEEauthorrefmark{1}
}

\IEEEauthorblockA{
\IEEEauthorrefmark{1}
\textit{Dept. of Computer and Data Sciences, Case Western Reserve University},
Cleveland, OH, USA\\
\{vxk285,jxl1646,aao62,xxl1337,sxx214,vxc204\}@case.edu
}

\IEEEauthorblockA{
\IEEEauthorrefmark{2}
\textit{Troy High School},
Fullerton, USA\\
800026128@fjuhsd.org
}

\IEEEauthorblockA{
\IEEEauthorrefmark{3}
\textit{Department of Molecular Medicine, Cleveland Clinic Lerner College of Medicine}\\
\textit{Case Western Reserve University},
Cleveland, OH, USA\\
blanked2@ccf.org
}
}

\maketitle

\begin{abstract}
As quantum computing scales toward practicality, future systems will shift from relying on a single monolithic device to using modular architectures that connect multiple QPUs. This transition is driven by the physical resources needed to support larger workloads and fault-tolerant computation. However, hardware technologies realize modularity through distinct communication models: superconducting platforms use real-time classical links and dynamic-circuit-based coordination, trapped-ion systems use photonic quantum interconnects for remote entanglement across modules, and neutral-atom architectures provide strong intra-core connectivity with proposed optical interconnects for inter-core communication. This heterogeneity makes communication-aware scheduling necessary for efficient execution in shared modular quantum cloud environments.

We present InterQ, a communication-aware scheduler for modular QPU architectures with heterogeneous communication models. InterQ jointly considers qubit capacity, qubit placement, parallel execution opportunities, and communication-driven dependencies across distributed subcircuits, while also enabling adaptive circuit cutting: the partitioning of jobs based on workload conditions to reduce makespan while balancing fidelity and communication overhead. The framework distinguishes between classical-link execution, where measurement and feedforward impose synchronization constraints, and quantum-link execution, where entanglement distribution and state transfer determine coordination cost. For neutral-atom platforms, InterQ further models intra-core and inter-core connectivity during qubit placement and scheduling. Using a unified framework to compare superconducting, trapped-ion, and neutral-atom modular systems, InterQ shows how communication models and scheduler-driven cutting decisions affect throughput, latency, and fidelity in scalable modular quantum computing. Across the evaluated workloads, InterQ exposes an architecture-dependent trade-off: neutral-atom modular QPUs achieve the highest fidelity, superconducting systems minimize runtime, and trapped-ion systems provide a balanced intermediate profile across fidelity and makespan.
\end{abstract}

\section{Introduction}
\label{sec:introduction}

Quantum computing is increasingly transitioning toward modular architectures in which multiple quantum processing units (QPUs) cooperate to execute workloads that exceed the practical limits of a single monolithic device. This shift is driven by both hardware scaling constraints and growing application demands. NISQ-era platforms continue to face fundamental limitations arising from noise, calibration overhead, constrained qubit connectivity, and the significant physical-qubit resources required to achieve fault-tolerant operation \cite{preskill2018quantum, roffe2019quantum}. Consequently, distributed and modular quantum computing has emerged as a viable systems-level paradigm across diverse hardware platforms rather than a platform-specific approach \cite{Barral_2025, Main_2025, Carrera_Vazquez_2024}.

Modularity changes scheduling from a pure packing problem into a communication-constrained coordination problem. In a shared quantum cloud, jobs compete not only for qubits and execution slots, but also for inter-module bandwidth, feed-forward latency, and entanglement-generation capacity. Once a circuit is partitioned across QPUs, the execution cost depends on the mechanism used to couple the resulting fragments. Purely local cutting can avoid immediate interconnect use, but it may incur severe subexperiment growth and expensive classical reconstruction: for example, wire cutting can scale as $\mathcal{O}(16^k)$ for $k$ cuts, while repeated two-qubit gate cutting can incur $\mathcal{O}(9^k)$ sampling overhead depending on the decomposition \cite{tang2021cutqc,Schmitt2025cuttingcircuits,qiskit-addon-cutting}. LOCC-based cutting can reduce wire-cutting overhead to $\mathcal{O}(4^k)$, but only by introducing real-time measurement, communication, and conditional-execution dependencies that can stall downstream fragments and reduce parallelism \cite{Piveteau_2024,brenner2023optimalwirecuttingclassical}. A scheduler that models only local occupancy and runtime can therefore make poor decisions by selecting cuts or placements that appear capacity-efficient but create excessive reconstruction work, synchronization delay, or communication bottlenecks.

These trade-offs are platform-dependent and become sharper in modular clouds where concurrent jobs share limited communication resources. Superconducting systems support LOCC-style execution through real-time classical communication, dynamic circuits, and measurement-conditioned feedforward \cite{Carrera_Vazquez_2024, PhysRevResearch.6.013235}. Trapped-ion systems use photonic interconnects with costs determined by probabilistic entanglement generation, retries, and link fidelity \cite{PhysRevLett.130.050803, Main_2025}, while neutral-atom systems combine strong local programmability with emerging optical interconnects \cite{li2024neutralatominterconnects}. Thus, interconnect resources must be scheduled carefully: aggressive quantum communication can serialize jobs on shared links, while excessive local or LOCC-based decomposition increases fragmentation and classical reconstruction cost. This motivates adaptive circuit cutting and communication-aware placement, where connectivity improves throughput only when used judiciously under dynamic workloads \cite{kulkarni2025quflex, 10821424}. Software support for modular execution and circuit knitting is also emerging through Qiskit-based toolchains \cite{Qiskit,qiskit-addon-cutting}.

Prior work addresses important parts of this space, but usually in isolation. Multi-programming and job-grouping systems improve throughput on shared devices by co-scheduling compatible jobs and optimizing qubit placement \cite{10.1145/3631525,niu2023enabling,10821381,guerreschi2018two,luo2025adaptive}. Circuit-cutting systems reduce width or connectivity requirements by decomposing a large circuit into smaller fragments, but incur sampling and reconstruction overheads that depend strongly on where and how the circuit is cut \cite{tang2021cutqc,Piveteau_2024,Perlin_2020,peng2020simulating,lowe2023fast,Schmitt2025cuttingcircuits,ufrecht2023cutting,Chen_2022,chen2023efficient,li2024efficient,qiskit-addon-cutting}. More recent work has begun combining cutting with runtime scheduling, including adaptive cut selection \cite{kulkarni2025quflex} and resource-constrained distributed execution \cite{10821424}. Related modular scheduling work has also shown that LO and LOCC execution semantics require explicit handling of causal dependencies between fragments \cite{kulkarni2026qumod}. However, a unified scheduler comparing classical-link and quantum-link communication modes remains missing.

InterQ addresses this gap by treating communication as a primary scheduling constraint rather than a post hoc placement correction. It distinguishes among three execution regimes: local operations (LO), in which fragments execute independently and correlations are reconstructed offline; local operations with classical communication (LOCC), in which mid-circuit measurement outcomes and feed-forward introduce causal dependencies across fragments; and quantum-link execution (QComm), in which coordination across QPUs is mediated by entanglement generation and Bell-pair consumption over shared interconnects \cite{Piveteau_2024,Carrera_Vazquez_2024,Main_2025,PhysRevLett.130.050803,li2024neutralatominterconnects}. InterQ incorporates these communication semantics directly into optimization decisions, jointly determining when to introduce cuts, how to place fragments across QPUs, and how to group them for concurrent execution.

The contributions of this paper are as follows:
\begin{itemize}
\item We formulate communication-aware scheduling for modular quantum systems as a constrained optimization problem over fragment generation, qubit placement, communication mode selection, and parallel execution, capturing capacity, precedence, and interconnect-feasibility constraints.
\item We present InterQ, a scheduler that distinguishes LO, LOCC, and quantum-link-based QComm execution and models their runtime effects, including LOCC synchronization, remote-operation generation, Bell-pair consumption, and shared-link contention.
\item We develop a constraint-aware methodology that combines adaptive circuit cutting with communication-sensitive grouping and placement, deciding when circuits should remain local, be partitioned, or use remote operations while balancing makespan, fidelity, and reconstruction overhead.
\item We provide a unified architecture-aware framework spanning superconducting, trapped-ion, and neutral-atom modular systems, enabling direct comparison of classical-link and quantum-link communication costs.
\item We evaluate InterQ on workloads from MQT Bench \cite{quetschlich2023mqtbench}, QUEKO \cite{tan2020queko}, RevLib \cite{wille2008revlib}, and random circuits, quantifying effects on throughput, waiting time, makespan, remote-operation demand, and distributed execution overhead.
\end{itemize}

The remainder of this paper is organized as follows. Section~\ref{sec:background} provides background on distributed execution models, hardware-specific modular architectures, and related scheduling systems. Subsequent sections formalize the InterQ problem, describe the scheduler design, and evaluate the resulting policies on representative workloads.

\section{Background}
\label{sec:background}

\subsection{Global Scheduling Context}
\label{subsec:global-scheduling}

We consider a modular quantum cloud comprising a set of modules $\mathcal{M} = \{1,\dots,M\}$, where each module $m \in \mathcal{M}$ is characterized by a qubit capacity $Q_m$, a local calibration profile $\Phi_m$, and a set of communication links to other modules. A job $J_i$ is defined by the tuple
\begin{equation}
J_i = (q_i, d_i, s_i, G_i, \chi_i),
\end{equation}
where $q_i$ denotes the logical qubit requirement, $d_i$ is a proxy for circuit depth, $s_i$ is the requested shot count, $G_i$ represents the circuit interaction graph or dependency structure, and $\chi_i$ specifies the admissible communication modalities for execution. For each job, the scheduler determines a partition $P_i$, a placement function $\pi_i : P_i \rightarrow \mathcal{M}$, and start times for the resulting fragments.

In contrast to single-QPU scheduling, the objective extends beyond minimizing queueing delay or maximizing instantaneous qubit utilization. Modular scheduling must additionally account for communication-induced delays, reconstruction overheads, and the cost of remote operations. These factors depend on how distributed execution is realized—whether through local execution, classical coordination, or quantum interconnects—and directly influence both system throughput and execution latency. This dependence motivates a communication-aware scheduling formulation.

\subsection{Distributed Execution Models}
\label{subsec:execution-models}

\subsubsection{Local Operations (LO)}
\label{subsubsec:lo}

In the LO setting, a cut severs a circuit into fragments that can be executed independently on separate devices or time slots, with nonlocal correlations reconstructed only after execution. Formally, a nonlocal channel $\mathcal{E}$ acting across a bipartition can be written as a quasi-probability decomposition
\begin{equation}
\mathcal{E} = \sum_{\alpha} w_{\alpha}
\bigl(\mathcal{E}^{(L)}_{\alpha} \otimes \mathcal{E}^{(R)}_{\alpha}\bigr),
\qquad
\kappa(\mathcal{E}) = \left(\sum_{\alpha} |w_{\alpha}|\right)^2,
\label{eq:qpd-lo}
\end{equation}
where the branch channels are implementable locally and the reconstruction cost is governed by the squared $\ell_1$ norm of the weights. For wire cutting, the effective sampling overhead typically scales as $\mathcal{O}(16^k)$ for $k$ cut wires, while a cut two-qubit gate can induce smaller but still multiplicative overheads depending on the gate decomposition \cite{tang2021cutqc,Piveteau_2024,Schmitt2025cuttingcircuits,qiskit-addon-cutting}.

From a scheduler's viewpoint, LO offers maximum placement freedom because fragments are independent once generated. The cost is paid offline through subexperiment proliferation and post-processing. This makes LO attractive when communication is unavailable or expensive, but increasingly unattractive as the number of cuts grows.

\subsubsection{Local Operations and Classical Communication (LOCC)}
\label{subsubsec:locc}

LOCC augments local fragment execution with real-time classical feed-forward. A cut wire can be implemented using upstream measurement, classical transmission of the outcome, and downstream conditional correction, yielding an exponential reduction in sampling overhead relative to purely local reconstruction \cite{Piveteau_2024,brenner2023optimalwirecuttingclassical}. For $k$ wire cuts, the effective scaling improves from $\mathcal{O}(16^k)$ to $\mathcal{O}(4^k)$ under idealized assumptions.

This reduction in reconstruction cost is offset by on-line precedence constraints. If fragment $u$ produces classical data required by fragment $v$, then the schedule must satisfy
\begin{equation}
t(v) \ge t(u) + p(u) + \delta^{\mathrm{meas}}_{u} + \delta^{\mathrm{tx}}_{u,v} + \delta^{\mathrm{ctrl}}_{v},
\label{eq:locc-precedence}
\end{equation}
where $p(u)$ is the fragment runtime and the additional terms denote measurement, transmission, and feed-forward control latency. LOCC therefore transforms what is an embarrassingly parallel fragment set in LO mode into a causally constrained pipeline. This distinction is critical in shared clouds, because the reduced sampling overhead of LOCC can be offset by increased waiting time if the scheduler cannot hide these dependencies through other concurrent work.

\begin{figure*}[t]
    \centering
    \includegraphics[width=\textwidth]{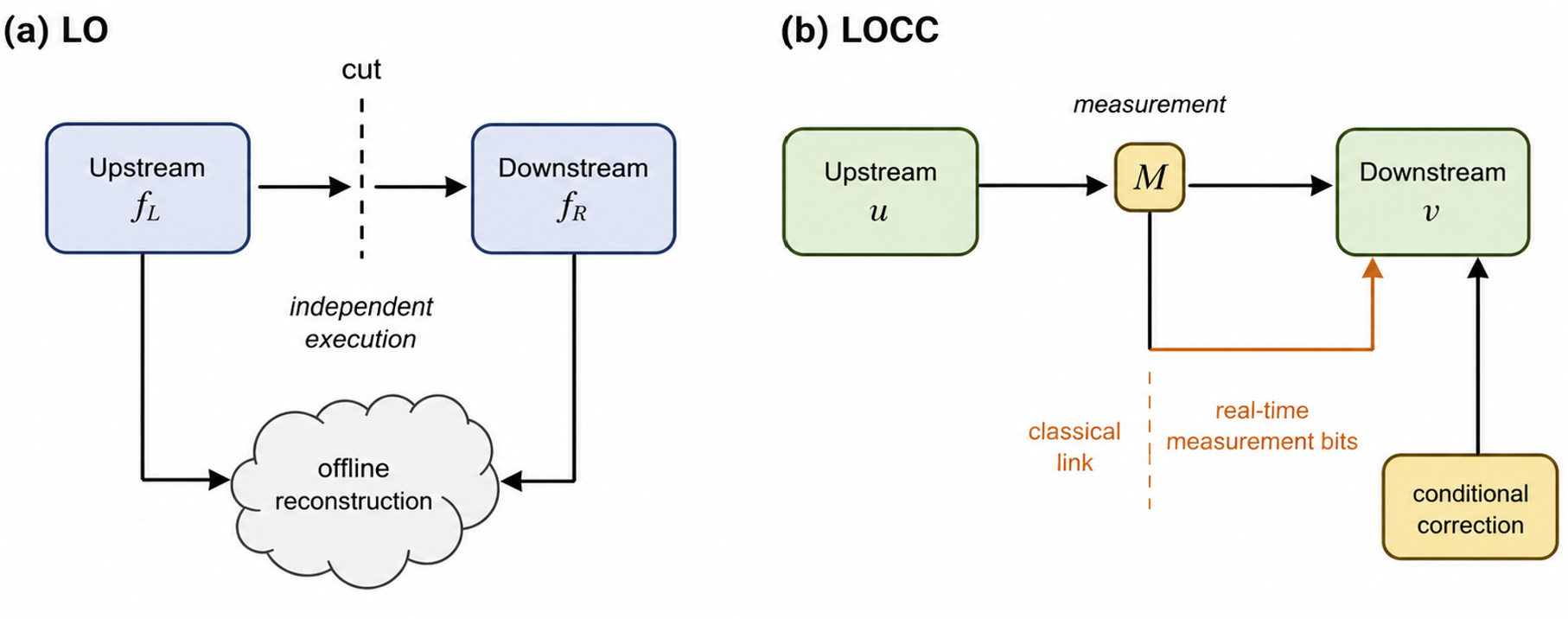}
    \caption{Execution semantics of circuit cutting. In LO, cut fragments execute independently and are stitched through offline reconstruction. In LOCC, upstream measurement outcomes are transmitted over a classical link and trigger downstream conditional operations, creating explicit precedence between fragments.}
    \label{fig:interq_lo_locc_cutting}
\end{figure*}

Figure~\ref{fig:interq_lo_locc_cutting} contrasts these two cutting semantics at the execution level.

\subsubsection{Quantum Communication (QComm)}
\label{subsubsec:qcomm}

In the QComm execution model, inter-module interactions are mediated by entanglement distribution rather than by offline reconstruction or purely classical feed-forward. For a remote operation $g$ executed between modules $a$ and $b$, a first-order scheduler-facing cost can be expressed as
\begin{equation}
c^{\mathrm{q}}(g)
= \frac{T_{\mathrm{pair}}(a,b)}{p_{\mathrm{succ}}(a,b)}
+ T_{\mathrm{bell}}(g)
+ T_{\mathrm{corr}}(g),
\label{eq:qcomm-cost}
\end{equation}
where $T_{\mathrm{pair}}(a,b)$ denotes the Bell-pair generation time between modules $a$ and $b$, $p_{\mathrm{succ}}(a,b)$ denotes the corresponding entanglement-generation success probability, and $T_{\mathrm{bell}}(g)$ and $T_{\mathrm{corr}}(g)$ capture local Bell-measurement and correction overheads, respectively.

The fidelity of a remote operation can be modeled multiplicatively as
\begin{equation}
F^{\mathrm{q}}(g) \approx F_{\mathrm{pair}}(a,b) F_{\mathrm{local}}(g) F_{\mathrm{meas}}(g),
\label{eq:qcomm-fidelity}
\end{equation}
where the constituent terms capture the quality of the distributed Bell pair, the local operation fidelity, and the measurement fidelity. This model highlights that remote execution consumes both execution-time and fidelity budgets.

From a scheduling perspective, quantum links must therefore be treated as explicit shared resources. When multiple jobs require remote entanglement concurrently, entanglement generation and remote-gate service capacity can become system bottlenecks. Unlike LOCC execution, where classical communication primarily introduces precedence and synchronization delays, QComm execution introduces contention for stochastic entanglement creation and link-service availability. Consequently, link-aware job selection, grouping, and placement are central to maintaining throughput in modular quantum systems.

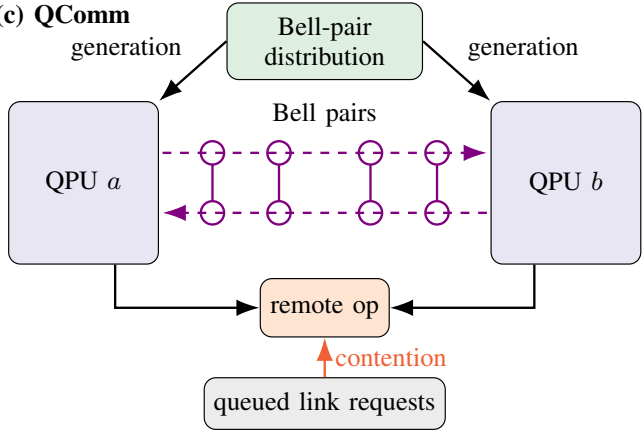
\begin{figure}[t]
    \centering
    \resizebox{0.98\columnwidth}{!}{%
    \begin{tikzpicture}[
        font=\small,
        >=Latex,
        flow/.style={-{Latex[length=2.8mm,width=2mm]}, thick},
        req/.style={-{Latex[length=2.8mm,width=2mm]}, thick, RedOrange},
        qpair/.style={-{Latex[length=2.8mm,width=2mm]}, thick, dashed, violet}
    ]
        \node[font=\small\bfseries] at (-0.25,2.0) {(c) QComm};
        \node[draw, rounded corners, fill=Blue!8, minimum width=1.8cm, minimum height=1.95cm] (qa) at (0,0) {QPU $a$};
        \node[draw, rounded corners, fill=Blue!8, minimum width=1.8cm, minimum height=1.95cm] (qb) at (5.8,0) {QPU $b$};
        \node[draw, rounded corners, fill=Green!12, minimum width=2.35cm, minimum height=0.92cm, align=center] (src) at (2.9,1.7) {Bell-pair\\distribution};
        \draw[flow] (src.west) -- node[above left] {generation} (qa.north east);
        \draw[flow] (src.east) -- node[above right] {generation} (qb.north west);
        \foreach \x in {1.55,2.35,3.45,4.25} {
            \draw[thick, violet] (\x,0.36) circle (0.14);
            \draw[thick, violet] (\x,-0.36) circle (0.14);
            \draw[thick, violet] (\x,0.22) -- (\x,-0.22);
        }
        \draw[qpair] (0.95,0.36) -- (4.85,0.36);
        \draw[qpair] (4.85,-0.36) -- (0.95,-0.36);
        \node at (2.9,0.82) {Bell pairs};
        \node[draw, rounded corners, fill=Orange!18, minimum width=1.55cm, minimum height=0.7cm] (gate) at (2.9,-1.5) {remote op};
        \draw[flow] ($(qa.south)+(0.38,0)$) -- ++(0,-0.45) |- (gate.west);
        \draw[flow] ($(qb.south)+(-0.38,0)$) -- ++(0,-0.45) |- (gate.east);
        \node[draw, rounded corners, fill=gray!15, minimum width=2.6cm, minimum height=0.62cm] (queue) at (2.9,-2.65) {queued link requests};
        \draw[req] (queue.north) -- node[right] {contention} (gate.south);
    \end{tikzpicture}%
    }
    \caption{QComm abstraction used by InterQ. Remote execution across modules consumes Bell-pair resources managed by a shared Bell-pair distribution service, so latency depends on Bell-pair generation, success probability, and contention from concurrent link requests.}
    \label{fig:interq_qcomm_bellpair}
\end{figure}

Figure~\ref{fig:interq_qcomm_bellpair} illustrates the QComm abstraction used by InterQ, where remote execution consumes Bell-pair resources on a shared link service.

\subsection{Hardware-Specific Modular Architectures}
\label{subsec:hardware-architectures}

\subsubsection{Superconducting Modular Systems}
\label{subsubsec:superconducting}

Superconducting platforms are among the most mature quantum cloud systems, but scaling monolithic devices remains challenging due to planar layouts, wiring constraints, and calibration overhead. Recent demonstrations combine multiple processors using real-time classical communication and dynamic circuits, enabling cross-processor coordination via measurement and feed-forward \cite{Carrera_Vazquez_2024}. Measurement-based virtual-gate constructions further allow nonlocal interactions to be emulated without direct quantum links \cite{Mitarai:2019vae,PhysRevResearch.6.013235,PhysRevLett.130.110601}. From a scheduling perspective, this architecture aligns naturally with LOCC-style execution, where performance is dominated by synchronization delays and conditional-operation latency.

\subsubsection{Trapped-Ion Modular Systems}
\label{subsubsec:trappedion}

Trapped-ion systems provide high-fidelity local operations and support modular scaling through photonic interconnects. Experimental results demonstrate remote entanglement over long distances and distributed execution across modules \cite{PhysRevLett.130.050803,Main_2025}. In this setting, execution is more naturally modeled using QComm, where entanglement generation, success probability, and link fidelity determine remote operation cost. Scheduling must therefore account for both qubit availability and the stochastic behavior of entanglement resources.

\subsubsection{Neutral-Atom Modular Systems}
\label{subsubsec:neutralatom}

Neutral-atom platforms exhibit a hierarchical structure: strong local interactions within arrays and emerging interconnect mechanisms for larger-scale modularity \cite{li2024neutralatominterconnects}. As a result, intra-core and inter-core operations have fundamentally different costs. This motivates a hierarchical scheduling model in which placement decisions distinguish between local execution and cross-module communication, rather than treating the system as a uniform resource pool.

\subsection{Related Scheduling and Cutting Systems}
\label{subsec:related-work}

Several prior systems motivate InterQ. QuCloud+ and related multi-programming approaches improve parallel execution on shared quantum devices through workload grouping and fidelity-aware mapping~\cite{10.1145/3631525,niu2023enabling}. QGroup optimizes parallel grouping using dynamic programming~\cite{10821381}, while other work studies circuit timing, resource allocation, adaptive job management, and VQA scheduling~\cite{guerreschi2018two,luo2025adaptive,10764550,li2025qusplit}. Workload-aware adaptive cutting shows that partitioning must account for queue state and placement constraints because cutting overhead can offset parallelization benefits~\cite{kulkarni2025quflex}, and resource-constrained distributed cut scheduling has also been studied~\cite{10821424}.

These systems typically address scheduling or partitioning under a fixed execution model. In contrast, modular quantum execution requires these decisions to be handled jointly because partitioning changes schedulable units, LOCC introduces synchronization dependencies, and quantum-link execution adds communication-resource constraints. InterQ builds on this direction by comparing heterogeneous interconnect models under classical-link and quantum-link execution assumptions.


\section{InterQ Scheduler}
\label{sec:interq-scheduler}

InterQ is a communication-aware scheduler for modular quantum execution that integrates QPU assignment, circuit partitioning, parallel grouping, and communication-resource management. It first maps jobs to available QPUs and identifies circuits that exceed single-device capacity as partitioning candidates. These candidates are expanded according to the target execution model, including LO, LOCC, or QComm. InterQ then groups compatible circuits for parallel execution while balancing makespan, fidelity, synchronization delay, and communication contention. A partitioning decision is accepted only when it preserves feasibility and improves the communication-aware scheduling objective.

\subsection{Communication-Aware Feasibility and Cost}
\label{subsec:interq-cost}

For a candidate group $g$ assigned to module $m$, InterQ evaluates capacity, precedence, and communication feasibility jointly. Let $q_j$ denote the qubit demand of job or fragment $j$, let $\sigma(j) \in \{\text{flat},\text{up},\text{down},\text{remote}\}$ denote its execution stage, and let $\rho(j)$ denote the parent job from which $j$ is derived when cutting is applied. A group is feasible on module $m$ only if its aggregate qubit demand satisfies
\begin{equation}
\sum_{j \in g} q_j \le Q_m ,
\label{eq:interq-feasible-capacity}
\end{equation}
and the corresponding mode-specific precedence and communication constraints are satisfied.

For LOCC execution, upstream and downstream fragments generated from the same parent cut cannot be placed in the same parallel group. InterQ therefore enforces
\begin{equation}
\rho(j)=\rho(k) \neq \bot
\;\Longrightarrow\;
\sigma(j)=\sigma(k),
\quad \forall j,k \in g .
\label{eq:interq-feasible-locc}
\end{equation}
This constraint prevents fragments with causal ordering dependencies from being scheduled as if they were independent.

For quantum-link execution, the remote-operation demand induced on each incident communication edge must not exceed the available link budget:
\begin{equation}
\sum_{j \in g} b_e(j) \le B_e,
\qquad \forall e \in \delta(m),
\label{eq:interq-feasible-qcomm}
\end{equation}
where $b_e(j)$ denotes the Bell-pair or remote-link demand generated by $j$ on edge $e$, $B_e$ is the available communication budget on that edge, and $\delta(m)$ is the set of links adjacent to module $m$.

InterQ assigns each feasible grouping and placement a communication-aware cost:
\begin{equation}
d_{\mathrm{InterQ}}(g,m) =
\alpha a(g) + \beta b(g,m) + \gamma c(g,m) + \eta h(g),
\label{eq:interq-group-cost}
\end{equation}
where
\begin{align}
a(g) &= \frac{\max_{j \in g} T_j}{\min_{j \in g} T_j} - 1,
\label{eq:interq-a-term} \\
b(g,m) &= \sum_{j \in g} \Delta_j^{\mathrm{sync}}(m),
\label{eq:interq-b-term} \\
c(g,m) &= \sum_{j \in g} \Omega_j^{\mathrm{comm}}(m),
\label{eq:interq-c-term} \\
h(g) &= \sum_{j \in g} \Omega_j^{\mathrm{cut}} .
\label{eq:interq-h-term}
\end{align}
The term $a(g)$ penalizes runtime imbalance within a parallel group, $b(g,m)$ captures synchronization delay and precedence slack, $c(g,m)$ models communication pressure on the selected module and its incident links, and $h(g)$ accounts for cut-induced sampling overhead. The weights $\alpha$, $\beta$, $\gamma$, and $\eta$ control the relative importance of these terms.

Given a schedule $\mathcal{S}$ consisting of groups mapped to modules, InterQ minimizes the aggregate scheduling objective
\begin{equation}
\mathcal{C}_{\mathrm{InterQ}}(\mathcal{S})
=
\sum_{(g,m) \in \mathcal{S}} d_{\mathrm{InterQ}}(g,m).
\label{eq:interq-objective}
\end{equation}
This objective provides a unified scheduler-facing criterion for comparing local execution, classical-link execution, and quantum-link execution without requiring separate mode-specific heuristics.

\begin{algorithm}[htbp]
\caption{InterQ Scheduling Algorithm}
\label{alg:interq_main}
\begin{algorithmic}[1]
    \Statex \textbf{Input:} job queue $\mathcal{J}$, modules $\mathcal{M}$, communication graph $\mathcal{L}$
    \Statex \textbf{Output:} final schedule $S_{\mathrm{final}}$
    \State $\mathcal{J}_{\mathrm{current}} \gets \mathcal{J}$
    \Repeat
        \State $improved \gets \textbf{false}$
        \State $\mathcal{G} \gets \textsc{PartitionInterQ}(\mathcal{J}_{\mathrm{current}}, \mathcal{M}, \mathcal{L})$
        \State $S_{\mathrm{initial}} \gets \textsc{MapGroups}(\mathcal{G}, \mathcal{M}, \mathcal{L})$
        \State $Z_{\mathrm{initial}} \gets \textsc{Objective}(S_{\mathrm{initial}})$
        \For{each job $j$ in $S_{\mathrm{initial}}$}
            \State $\mathcal{C} \gets \textsc{Intercomm\_modes}(j, \mathcal{M}, \mathcal{L})$
            \For{each candidate fragment set $F \in \mathcal{C}$}
                \If{$F$ violates capacity, precedence, or link budgets}
                    \State \textbf{continue}
                \EndIf
                \State $\mathcal{J}_{\mathrm{cand}} \gets (\mathcal{J}_{\mathrm{current}} \setminus \{j\}) \cup F$
                \State $\mathcal{G}_{\mathrm{cand}} \gets \textsc{PartitionInterQ}(\mathcal{J}_{\mathrm{cand}}, \mathcal{M}, \mathcal{L})$
                \State $S_{\mathrm{cand}} \gets \textsc{MapGroups}(\mathcal{G}_{\mathrm{cand}}, \mathcal{M}, \mathcal{L})$
                \State $Z_{\mathrm{cand}} \gets \textsc{Objective}(S_{\mathrm{cand}})$
                \If{$Z_{\mathrm{cand}} < Z_{\mathrm{initial}}$}
                    \State $\mathcal{J}_{\mathrm{current}} \gets \mathcal{J}_{\mathrm{cand}}$
                    \State $improved \gets \textbf{true}$
                    \State \textbf{break}
                \EndIf
            \EndFor
            \If{$improved$}
                \State \textbf{break}
            \EndIf
        \EndFor
    \Until{not $improved$}
    \State $\mathcal{G}_{\mathrm{final}} \gets \textsc{PartitionInterQ}(\mathcal{J}_{\mathrm{current}}, \mathcal{M}, \mathcal{L})$
    \State $S_{\mathrm{final}} \gets \textsc{MapGroups}(\mathcal{G}_{\mathrm{final}}, \mathcal{M}, \mathcal{L})$
    \State \Return $S_{\mathrm{final}}$
\end{algorithmic}
\end{algorithm}

\subsection{Scheduler Overview}
\label{subsec:interq-overview}

The overall InterQ procedure is summarized in Algorithm~\ref{alg:interq_main}. Starting from the active job queue, InterQ iteratively forms communication-feasible groups, maps them to available modules, and constructs an initial modular schedule. It then evaluates whether selected scheduled jobs should be repartitioned into LO, LOCC, or QComm fragments according to the current workload state and available hardware resources. A candidate transformation is accepted only when the resulting schedule remains feasible under Eqs.~(\ref{eq:interq-feasible-capacity})--(\ref{eq:interq-feasible-qcomm}) and reduces the objective in Eq.~(\ref{eq:interq-objective}).

\subsection{Architecture-Aware Partitioning}
\label{subsec:interq-expansion}

Algorithm~\ref{alg:interq_expand} summarizes the architecture-aware partitioning stage. InterQ differs from prior schedulers by expanding jobs according to the execution modes supported by the target modular architecture. For LO execution, fragments are treated as independent units with offline reconstruction metadata. For LOCC execution, each cut introduces upstream and downstream fragments with precedence constraints and classical-delay annotations. For QComm execution, fragments are annotated with remote-gate counts, Bell-pair demand, and link-occupancy estimates. Candidate expansions that exceed the cut budget or communication budget are pruned before full schedule re-evaluation.

\begin{algorithm}[htbp]
\caption{Architecture-aware Partitioning}
\label{alg:interq_expand}
\begin{algorithmic}[1]
    \Function{Intercomm\_modes}{job $j$, modules $\mathcal{M}$, links $\mathcal{L}$}
        \State $\mathcal{C} \gets \emptyset$
        \ForAll{mode $\mu \in \chi(j)$}
            \State $(\Pi, \mathrm{meta}) \gets \textsc{FindPartition}(j, \mu)$
            \If{$\Pi = \emptyset$}
                \State \textbf{continue}
            \EndIf
            \State $F \gets \emptyset$
            \If{$\mu = \text{LO}$}
                \ForAll{fragment $f \in \Pi$}
                    \State $\sigma_f \gets \text{flat}$
                    \State $\Omega_f^{\mathrm{cut}} \gets \textsc{CutCost}(f)$
                    \State $F \gets F \cup \{f\}$
                \EndFor
            \ElsIf{$\mu = \text{LOCC}$}
                \ForAll{cut pair $(u,v)$ induced by $\Pi$}
                    \State $\sigma_u \gets \text{upstream}$; $\sigma_v \gets \text{downstream}$
                    \State $E_{\mathrm{prec}} \gets E_{\mathrm{prec}} \cup \{u \rightarrow v\}$
                    \State $\Delta_{u,v} \gets \delta^{\mathrm{meas}} + \delta^{\mathrm{tx}} + \delta^{\mathrm{ctrl}}$
                    \State $F \gets F \cup \{u,v\}$
                \EndFor
            \Else
                \ForAll{fragment $f \in \Pi$}
                    \State $r_f \gets \textsc{RemoteOps}(f)$
                    \State $b_f \gets \textsc{BellPairs}(f)$
                    \State $\Omega_f^{\mathrm{comm}} \gets \textsc{LinkCost}(f)$
                    \State $\sigma_f \gets \text{remote}$
                    \State $F \gets F \cup \{f\}$
                \EndFor
            \EndIf
            \If{$\sum_{f \in F} \Omega_f^{\mathrm{cut}} > \Theta_{\mathrm{cut}}$}
                \State \textbf{continue}
            \EndIf
            \If{$\sum_{f \in F} \Omega_f^{\mathrm{comm}} > \Theta_{\mathrm{comm}}$}
                \State \textbf{continue}
            \EndIf
            \State $\mathcal{C} \gets \mathcal{C} \cup \{F\}$
        \EndFor
        \State \Return $\mathcal{C}$
    \EndFunction
\end{algorithmic}
\end{algorithm}

\subsection{Grouping and Placement Under Communication Constraints}
\label{subsec:interq-grouping}

Following partitioning, InterQ forms parallel groups that satisfy both runtime-compatibility and communication-feasibility requirements. This stage applies execution-model-specific grouping rules: LO fragments may be co-scheduled whenever module capacity permits, LOCC fragments must preserve causal stage ordering, and QComm fragments must not oversubscribe the communication links associated with the selected placement. Algorithm~\ref{alg:interq_grouping} performs this selection greedily using the cost function in Eq.~(\ref{eq:interq-group-cost}). The same scoring function can also be incorporated into dynamic-programming or look-ahead grouping strategies when stronger optimization is desired.

\begin{algorithm}[htbp]
\caption{InterQ Grouping and Placement}
\label{alg:interq_grouping}
\begin{algorithmic}[1]
    \Function{PartitionInterQ}{$\mathcal{J}, \mathcal{M}, \mathcal{L}$}
        \State $\mathcal{U} \gets \textsc{SortByRuntime}(\mathcal{J})$
        \State $\mathcal{G} \gets \emptyset$
        \While{$\mathcal{U} \neq \emptyset$}
            \State $g^{\star} \gets \emptyset$
            \State $(m^{\star}, score^{\star}) \gets (\bot, \infty)$
            \ForAll{module $m \in \mathcal{M}$}
                \State $g \gets \emptyset$
                \State $Q_{\mathrm{used}} \gets 0$
                \ForAll{$j \in \mathcal{U}$}
                    \If{$Q_{\mathrm{used}} + q_j > Q_m$}
                        \State \textbf{continue}
                    \EndIf
                    \If{$\textsc{StageConflict}(j, g)$}
                        \State \textbf{continue}
                    \EndIf
                    \If{$\textsc{LinkBudget}(j, g, \delta(m))$}
                        \State \textbf{continue}
                    \EndIf
                    \State $g \gets g \cup \{j\}$
                    \State $Q_{\mathrm{used}} \gets Q_{\mathrm{used}} + q_j$
                \EndFor
                \If{$g \neq \emptyset$}
                    \State $score \gets d_{\mathrm{InterQ}}(g,m)$
                    \If{$score < score^{\star}$}
                        \State $(g^{\star}, m^{\star}, score^{\star}) \gets (g, m, score)$
                    \EndIf
                \EndIf
            \EndFor
            \State $\pi(g^{\star}) \gets m^{\star}$
            \State $\mathcal{G} \gets \mathcal{G} \cup \{(g^{\star}, m^{\star})\}$
            \State $\mathcal{U} \gets \mathcal{U} \setminus g^{\star}$
        \EndWhile
        \State \Return $\mathcal{G}$
    \EndFunction
    \Statex
    \Function{MapGroups}{$\mathcal{G}, \mathcal{M}, \mathcal{L}$}
        \State $S \gets \emptyset$
        \ForAll{$(g,m) \in \mathcal{G}$ in nondecreasing queue order}
            \State $t^{\star} \gets \textsc{EarliestStart}(g, m, S)$
            \State $S \gets \textsc{ReserveLinks}(S, g, m, t^{\star})$
            \State $S \gets \textsc{AddWithPrecedence}(S, g, m, t^{\star})$
        \EndFor
        \State \Return $S$
    \EndFunction
\end{algorithmic}
\end{algorithm}

Together, Algorithms~\ref{alg:interq_main}--\ref{alg:interq_grouping} define InterQ as a closed-loop scheduler: candidate jobs are repartitioned only when doing so improves the communication-aware objective, and grouping decisions explicitly reflect whether the architecture realizes distributed execution through offline reconstruction, classical synchronization, or quantum interconnects. The formulation is modular in the sense that the LO, LOCC, and QComm branches can be analyzed independently or composed within a unified scheduling framework.

\section{Evaluation}
\label{sec:evaluation}




InterQ is implemented in a SimPy-based discrete-event
simulation framework~\cite{10.7717/peerj-cs.103} with a job queue, modular
backend profiles, and a scheduler that tracks QPU occupancy and
communication resources over time. Circuit construction and
partitioning are aligned with Qiskit-based tooling~\cite{Qiskit,qiskit-addon-cutting}.
We evaluate three hardware-inspired modular architectures:
superconducting IBM-style LOCC execution, trapped-ion IonQ-style
QComm execution, and neutral-atom Atomic QComm execution. In QComm
mode, the simulator instantiates a mixed Aria/Forte resource pool
when the provider profile is set to IonQ, enabling comparison of
scheduling behavior under classical-link and quantum-link assumptions
within the same simulation environment.

The workloads combine synthetic circuits with benchmarks
from MQT Bench, QUEKO, and RevLib~\cite{quetschlich2023mqtbench,tan2020queko,wille2008revlib}.
The MQT Bench workloads include representative circuits such as
Real Amplitudes, QFT, EfficientSU2, Deutsch--Jozsa, VQE,
GHZ-state preparation, Amplitude Estimation, and TwoLocal.
Smaller workloads illustrate schedule structure and communication
dependencies, while larger workloads quantify partitioning behavior
and induced communication demand. We report workload width/depth
distributions, execution traces, queue evolution, and QComm-specific
metrics, including partition count, estimated remote operations, and
teleportation-induced resource overhead.
\subsection{Experimental Setup and Metrics}
\label{subsec:eval-setup}

For IBM-style experiments, we model modular superconducting execution with classical feed-forward and synchronization-sensitive scheduling, consistent with recent demonstrations of cross-processor classical coordination and virtual inter-processor gate construction~\cite{Carrera_Vazquez_2024,PhysRevResearch.6.013235}. For IonQ-style experiments, we use a trapped-ion-inspired QComm profile, where communication cost is dominated by remote entanglement or state-transfer demand~\cite{Main_2025,PhysRevLett.130.050803}.

Table~\ref{tab:eval_platform_config} summarizes the backend configurations and inter-module communication parameters used in the evaluation. For IBM-style LOCC, the classical-link latency, dynamic-circuit/control overhead, and cut limits are modeled as backend-informed scheduler-level simulation parameters for real-time classical coordination in the evaluation framework.

\begin{table*}[!t]
    \centering
    \caption{\small Platform configuration used in the IBM- LOCC, IonQ-QComm, and Atomic-QComm evaluation.}
    \label{tab:eval_platform_config}
    \footnotesize
    \setlength{\tabcolsep}{5pt}
    \renewcommand{\arraystretch}{1.12}
    \setlength{\arrayrulewidth}{0.5pt}
    \arrayrulecolor{black}

    \begin{tabular}{|p{3.0cm}|p{4.0cm}|p{3.9cm}|p{3.9cm}|}
        \hline
        \rowcolor{tableheaderdark}
        \textcolor{white}{\textbf{Parameter}} & \textcolor{white}{\textbf{IBM LOCC}} & \textcolor{white}{\textbf{IonQ QComm}} & \textcolor{white}{\textbf{Atomic QComm}} \\
        \hline
        \rowcolor{metricrowa}
        \textbf{Backend modules} & \texttt{ibm\_kawasaki}, \texttt{ibm\_kyiv}, \texttt{ibm\_sherbrooke}, \texttt{ibm\_brisbane} & \texttt{ionq\_aria}, \texttt{ionq\_forte} & \texttt{ac1000} \\
        \hline
        \rowcolor{metricrowb}
        \textbf{Per-module qubit capacity} & 127 qubits each & Aria: 25 qubits; Forte: 36 qubits & 112 physical qubits \\
        \hline
        \rowcolor{metricrowa}
        \textbf{Module count} & 4 & 6 & 4 \\
        \hline
        \rowcolor{metricrowb}
        \textbf{Classical / link latency} & $5.0 \times 10^5$ ns LOCC link latency & $2.0 \times 10^6$ ns QComm link latency & $3.0 \times 10^6$ ns QComm link latency \\
        \hline
        \rowcolor{metricrowa}
        \textbf{Control / feed-forward overhead} & $1.5 \times 10^6$ ns dynamic-circuit overhead & $2.0 \times 10^5$ ns classical feed-forward & $3.0 \times 10^5$ ns classical feed-forward \\
        \hline
        \rowcolor{metricrowb}
        \textbf{Remote-gate latency} & -- & $2.0 \times 10^5$ ns & $1.0 \times 10^5$ ns \\
        \hline
        \rowcolor{metricrowa}
        \textbf{Bell-pair generation rate} & -- & $5.0 \times 10^3$ Hz & $3.0 \times 10^3$ Hz \\
        \hline
        \rowcolor{metricrowb}
        \textbf{Bell-pair fidelity} & -- & 0.99 & 0.988 \\
        \hline
        \rowcolor{metricrowa}
        \textbf{Bell-pair TTL} & -- & $5.0 \times 10^8$ ns & $3.0 \times 10^8$ ns \\
        \hline
        \rowcolor{metricrowb}
        \textbf{Link parallelism} & Implicitly single classical dependency chain & 1 & 1 \\
        \hline
        \rowcolor{metricrowa}
        \textbf{Sampling factor} & LOCC effective wire-cut overhead model & 2.0 & 2.0 \\
        \hline
        \rowcolor{metricrowb}
        \textbf{Cut / overhead limit} & Max sampling overhead 16 & QComm module count 6 & QComm module count 4 \\
        \hline
    \end{tabular}
\end{table*}

We report average queue length, average queue time ($T_{\mathrm{wait}}$), average runtime ($T_{\mathrm{run}}$), average response time ($T_{\mathrm{total}}=T_{\mathrm{wait}}+T_{\mathrm{run}}$), workload changes, and fidelity-related metrics. LPST denotes the log probability of successful trial; values closer to $0$ indicate higher expected success probability. TRF denotes the throughput-ratio factor, TiRF denotes the time-reduction factor, and TiIF denotes the time-integrated fidelity factor. Higher values are preferred for TRF, TiRF, and TiIF, while lower values are preferred for queue length, queue time, runtime, wait time, and workload changes. The Subcircuits generated in IBM LOCC runs are named with JobID-U-subcircuit\# and JobID-D-subcircuit\# for upstream and corresponding downstream circuits. The partitioned circuits for IONQ and Atomic Qcomm modes are named with JOBID\_P and JOBID\_X to indicate the corresponding remote operations between the partitions. 

\subsection{IBM Adaptive LOCC vs. IonQ QComm}
\label{subsec:eval-ibm-locc-ionq-qcomm}

\begin{figure}[!t]
    \centering

    \begin{subfigure}[t]{\linewidth}
        \centering
        \includegraphics[width=\linewidth]{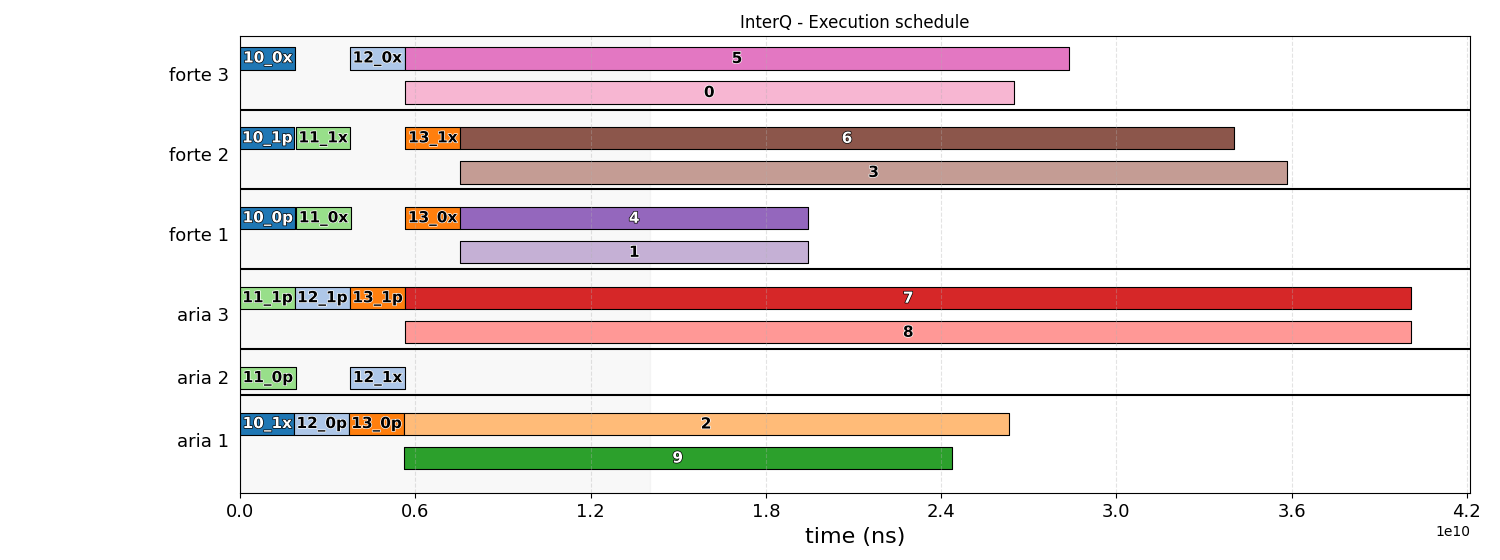}
        \caption{IonQ QComm schedule.}
        \label{fig:interq_eval_mbl_queko_ionq_schedule}
    \end{subfigure}

    \vspace{0.4em}

    \begin{subfigure}[t]{\linewidth}
        \centering
        \includegraphics[width=\linewidth]{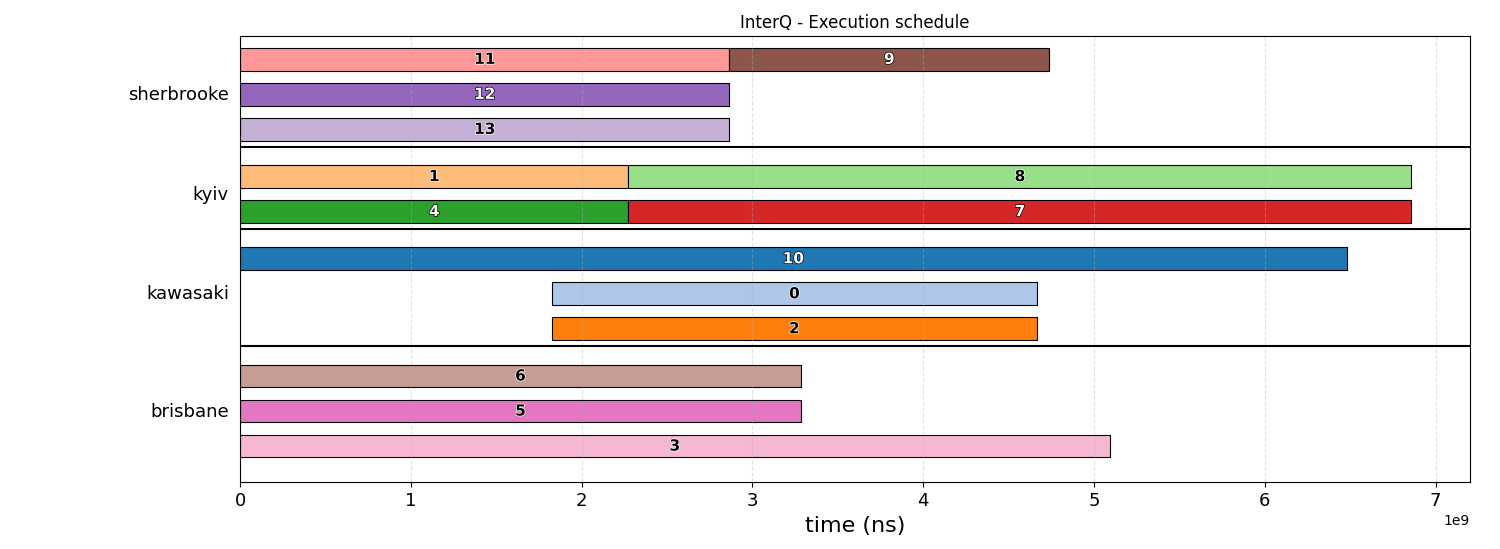}
        \caption{IBM LOCC schedule.}
        \label{fig:interq_eval_mbl_queko_ibm_schedule}
    \end{subfigure}

    \caption{InterQ schedules for the MQT and QUEKO benchmark workload under two communication models: (a) IonQ QComm, where scheduling is influenced by remote-operation demand and quantum-link availability, and (b) IBM adaptive LOCC mode, where no circuits are partitioned because all circuits fit within the available IBM QPU capacities.}
    \label{fig:interq_eval_mbl_queko_schedule_comparison}
\end{figure}

\begin{figure}[!t]
    \centering
    \includegraphics[width=\linewidth]{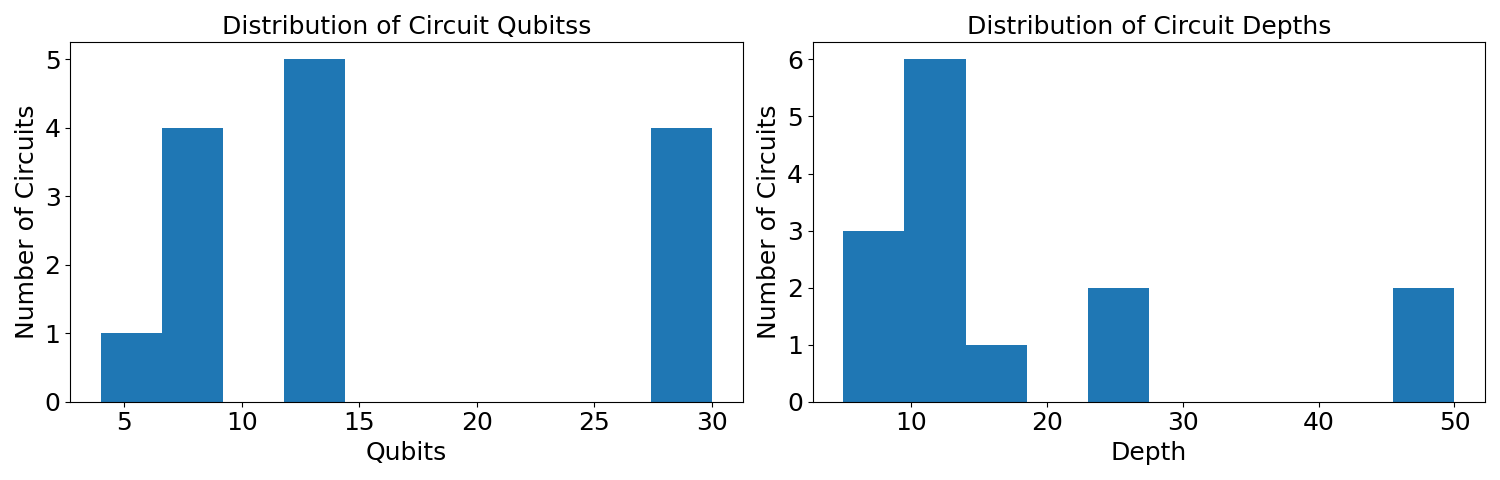}
    \caption{Workload distribution for the MQT and Queko benchmark set used in both the IBM LOCC and IonQ QComm runs.}
    \label{fig:interq_eval_mbl_queko_distribution}
\end{figure}

Table~\ref{tab:interq_ibm_ionq_copy} compares IBM LOCC and IonQ QComm scheduling on the MQT and QUEKO benchmark workload, whose circuit distribution is shown in Fig.~\ref{fig:interq_eval_mbl_queko_distribution}. Some circuits exceed the capacity of individual IonQ Aria QPUs and are partitioned under QComm, as shown in Fig.~\ref{fig:interq_eval_mbl_queko_schedule_comparison}. Partitioned QComm fragments are labeled with suffixes such as \texttt{\_p} and \texttt{\_x} to indicate Bell-state preparation and remote-operation fragments, which introduce communication delays in some cases. In contrast, the IBM configuration accommodates all circuits without partitioning, demonstrating InterQ's adaptive behavior. IBM LOCC achieves lower makespan, while IonQ QComm achieves better LPST; green cells mark the preferred metric values.

\begin{figure}[t]
\centering

\begin{subfigure}{\columnwidth}
    \centering
    \includegraphics[width=\columnwidth]{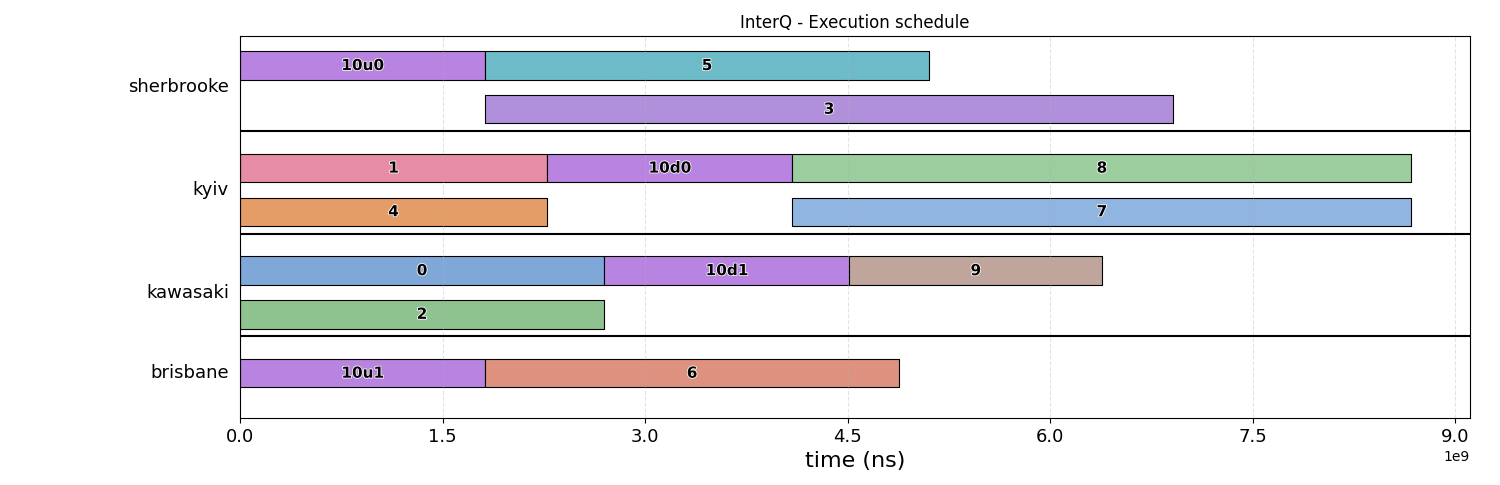}
    \caption{IBM LOCC}
\end{subfigure}

\vspace{4pt}

\begin{subfigure}{\columnwidth}
    \centering
    \includegraphics[width=\columnwidth]{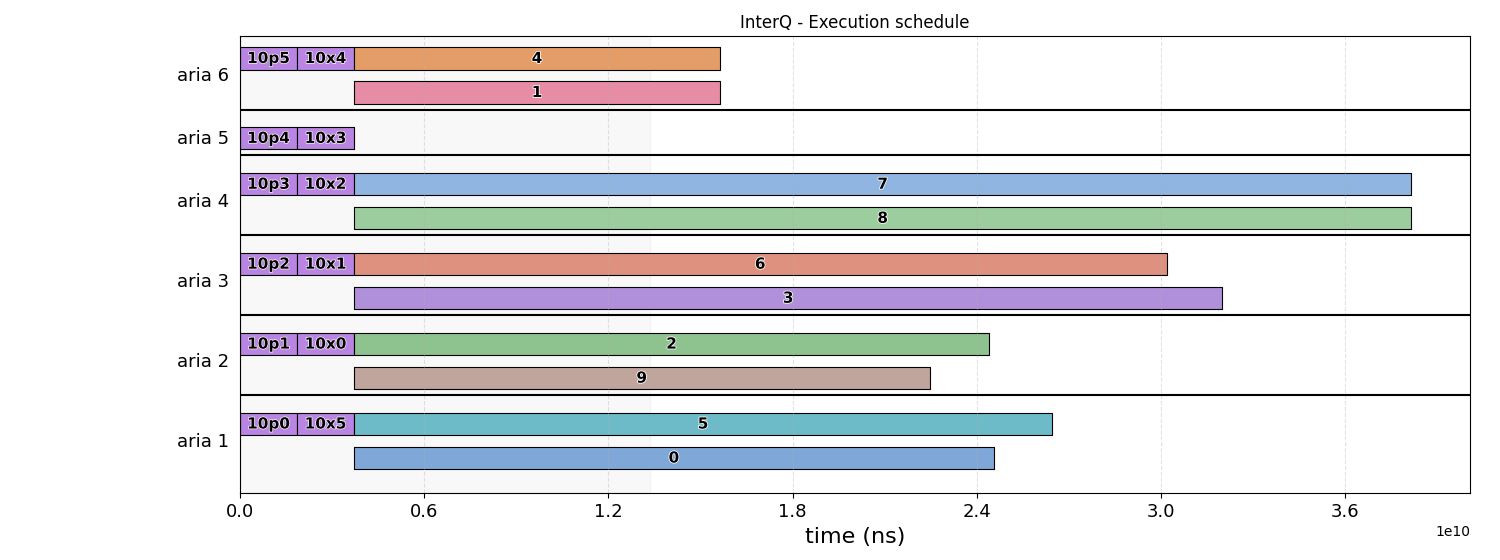}
    \caption{Atom QComm}
\end{subfigure}

\vspace{4pt}

\begin{subfigure}{\columnwidth}
    \centering
    \includegraphics[width=\columnwidth]{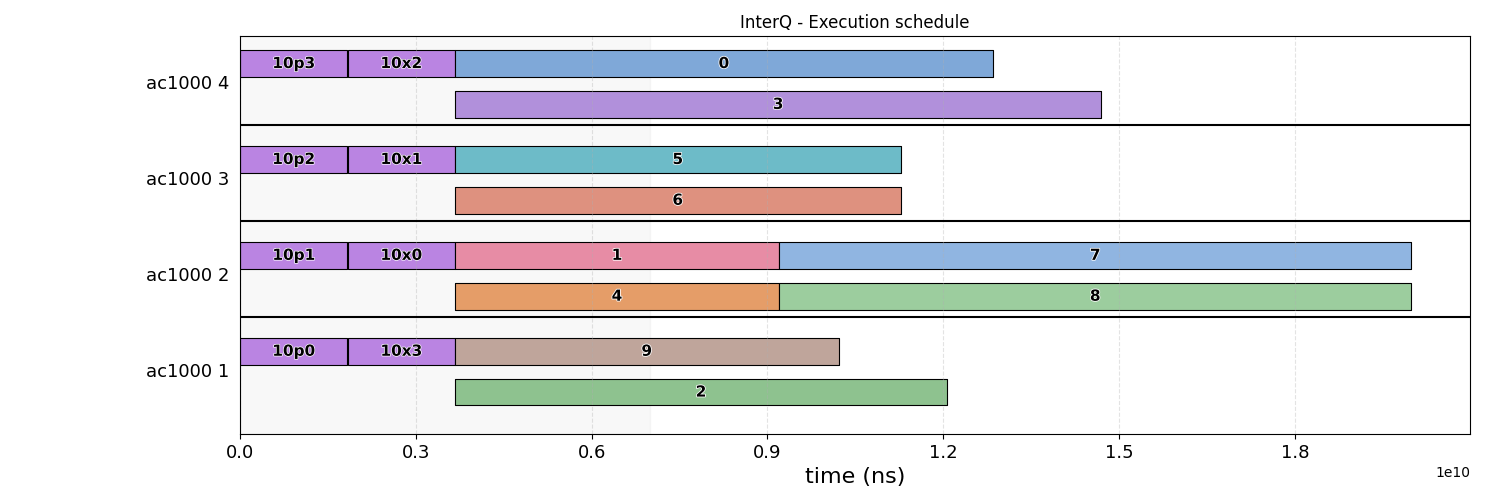}
    \caption{IONQ QComm}
\end{subfigure}

\caption{Execution schedule comparison of IBM LOCC, Atom QComm, and Ionq QComm architectures for the 11-circuit workload.}
\label{fig:ibm-ionq-atom-11c-schedules}

\end{figure}

\begin{table}[t]
\centering
\small
\setlength{\tabcolsep}{6pt}
\renewcommand{\arraystretch}{1.2}

\caption{Comparison of IBM LOCC, IonQ QComm, and Atom QComm for the 11-circuit workload. Green cells mark the preferred value for each metric.}
\label{tab:ibm-ionq-atom-11c-metrics}

\rowcolors{2}{gray!6}{white}
\begin{tabular}{|>{\raggedright\arraybackslash}p{3.2cm}|c|c|c|}
\hline
\rowcolor{blue!22}
\textbf{Metric} & \textbf{IBM} & \textbf{IonQ} & \textbf{Atom} \\
\hline
Avg. Run Time & \cellcolor{green!25}\textbf{2.9477} & 20.9625 & 7.5454 \\
\hline
Avg. TiIF & 0.9606 & \cellcolor{green!25}\textbf{1.0000} & 0.9357 \\
\hline
Avg. LPST & -1.9711 & -0.4696 & \cellcolor{green!25}\textbf{-0.2535} \\
\hline
\end{tabular}
\end{table}

\subsection{IBM LOCC vs. IBM Serial Round Robin}
\label{subsec:eval-ibm-serial-interq-15}

\begin{figure}[!t]
    \centering
    \includegraphics[width=\linewidth]{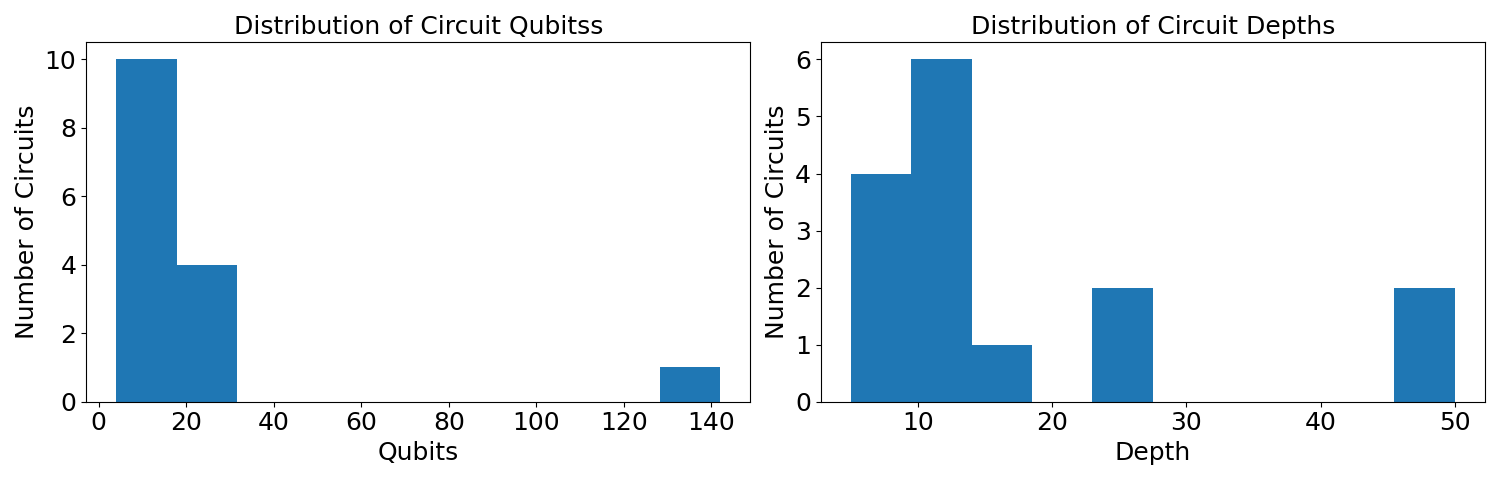}
    \caption{Circuit distribution for MBL \& Queko workload used in the IBM LOCC adaptive cutting vs serial Round Robin scheduling comparison.}
    \label{fig:ibm_serial_interq_15_distribution}
\end{figure}

This experiment compares LOCC-enabled InterQ with Serial Round Robin on a 15-circuit \texttt{MBL\_queko\_revlib} workload that includes a 142-qubit many-body localization (MBL) circuit exceeding the capacity of any individual 127-qubit backend. The circuit distribution is shown in Fig.~\ref{fig:ibm_serial_interq_15_distribution}.

\begin{figure}[!t]
    \centering

    \begin{subfigure}[t]{\linewidth}
        \centering
        \includegraphics[width=\linewidth]{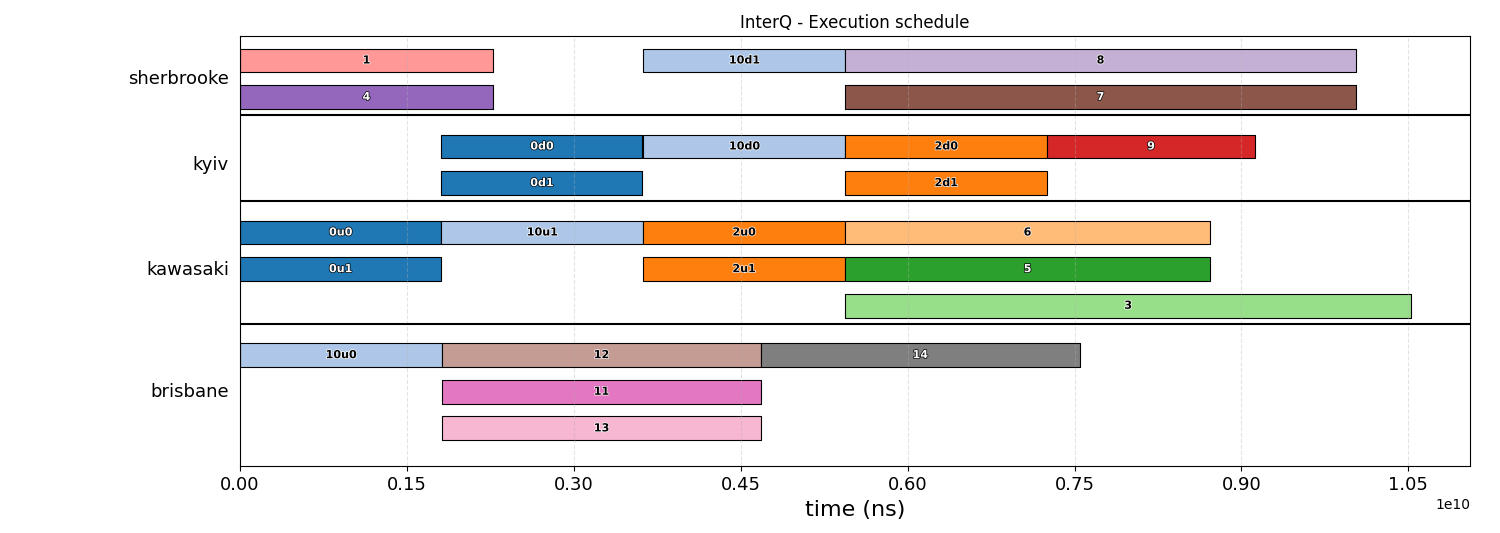}
        \caption{InterQ schedule with adaptive LOCC cutting.}
        \label{fig:ibm_serial_interq_15_interq_schedule}
    \end{subfigure}

    \vspace{0.4em}

    \begin{subfigure}[t]{\linewidth}
        \centering
        \includegraphics[width=\linewidth]{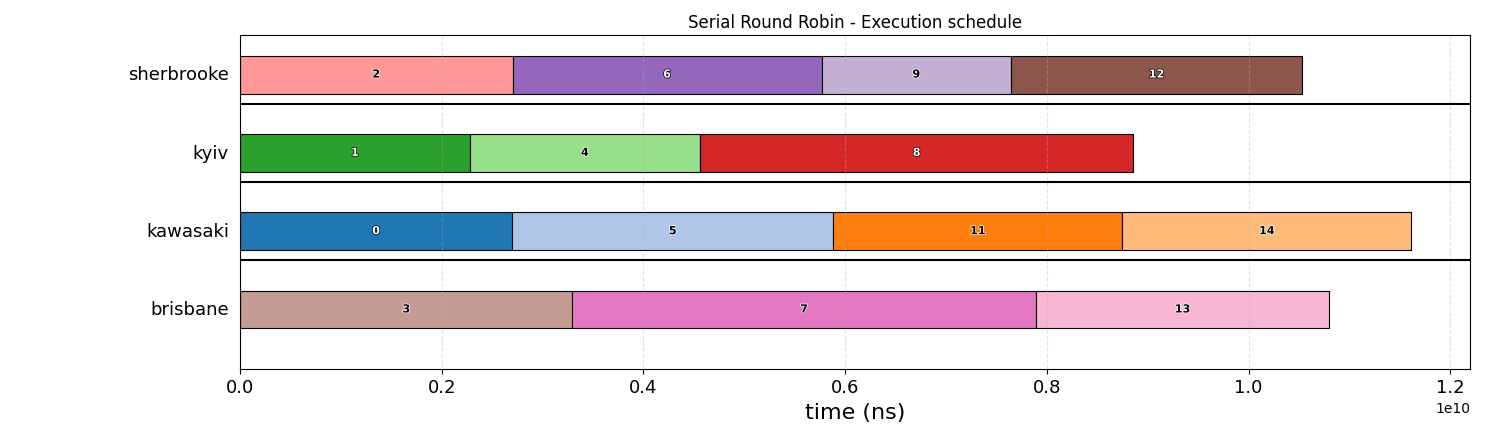}
        \caption{Serial Round Robin schedule.}
        \label{fig:ibm_serial_interq_15_serial_schedule}
    \end{subfigure}

    \caption{Schedule comparison for the 15-circuit IBM LOCC workload. (a) InterQ uses adaptive LOCC cutting to execute oversized circuits. (b) Serial Round Robin does not cut the 142-qubit circuit (Job 10), so it is absent from the schedule.}
    \label{fig:ibm_serial_interq_15_schedule_comparison}
\end{figure}

Figure~\ref{fig:ibm_serial_interq_15_schedule_comparison} compares the execution schedules, and Table~\ref{tab:ibm_serial_interq_15} reports the corresponding makespan and fidelity-related metrics. Serial Round Robin achieves better queue length, workload changes, TiIF, and LPST, whereas InterQ reduces queue time, runtime, and wait time while improving TRF and TiRF through adaptive LOCC partitioning. The 142-qubit circuit cannot be placed directly on any 127-qubit backend and is therefore omitted from the Serial Round Robin schedule; InterQ partitions it into LOCC-compatible upstream and downstream fragments and schedules them across the modular backend pool.

\begin{table}[!t]
    \centering
    \caption{\small IBM LOCC comparison on the updated 15-circuit \texttt{MBL\_queko\_revlib} queue with a 142-qubit circuit: Serial Round Robin versus IBM LOCC. Green cells indicate the better value for that metric. For PST, the value closer to $0$ is preferred.}
    \label{tab:ibm_serial_interq_15}
    \vspace{0.5ex}
    \setlength{\tabcolsep}{6pt}
    \renewcommand{\arraystretch}{1.18}
    \begin{tabular}{|>{\columncolor{metricrowa}}c|c|c|}
        \hline
        \rowcolor{tableheaderdark}
        \textcolor{white}{\textbf{Metric}} & \cellcolor{tableheaderibm}\textcolor{white}{\textbf{Serial RR}} & \cellcolor{tableheaderibm}\textcolor{white}{\textbf{InterQ}} \\
        \hline
        \rowcolor{metricrowa}
        \textbf{Average Queue Length} & \metricwinner{5.432} & 6.416 \\
        \hline
        \rowcolor{metricrowb}
        \textbf{Average Queue Time} & 3.431 & \metricwinner{2.972} \\
        \hline
        \rowcolor{metricrowa}
        \textbf{Average Run Time} & 2.787 & \metricwinner{2.581} \\
        \hline
        \rowcolor{metricrowb}
        \textbf{Average Wait Time} & 6.218 & \metricwinner{5.553} \\
        \hline
        \rowcolor{metricrowa}
        \textbf{Workload Changes} & \metricwinner{14} & 15 \\
        \hline
        \rowcolor{metricrowb}
        \textbf{TRF} & 1.000 & \metricwinner{2.316} \\
        \hline
        \rowcolor{metricrowa}
        \textbf{TiRF} & 1.000 & \metricwinner{1.227} \\
        \hline
        \rowcolor{metricrowb}
        \textbf{Average TiIF} & \metricwinner{1.000} & 0.968 \\
        \hline
        \rowcolor{metricrowa}
        \textbf{Average LPST} & \metricwinner{-0.973} & -1.913 \\
        \hline
    \end{tabular}
\end{table}

\begin{table}[!t]
    \centering
    \caption{\small InterQ comparison on combined MQT and Queko Benchark workload: IBM LOCC versus IonQ QComm. Green cells indicate the better value for that metric. For PST, the value closer to $0$ is preferred.}
    \label{tab:interq_ibm_ionq_copy}
    \vspace{0.5ex}
    \setlength{\tabcolsep}{6pt}
    \renewcommand{\arraystretch}{1.18}
    \begin{tabular}{|>{\columncolor{metricrowa}}c|c|c|}
        \hline
        \rowcolor{tableheaderdark}
        \textcolor{white}{\textbf{Metric}} & \cellcolor{tableheaderibm}\textcolor{white}{\textbf{IBM LOCC}} & \cellcolor{tableheaderionq}\textcolor{white}{\textbf{IonQ QComm}} \\
        \hline
        \rowcolor{metricrowa}
        \textbf{Average Queue Length} & \metricwinner{1.081} & 2.252 \\
        \hline
        \rowcolor{metricrowb}
        \textbf{Average Queue Time} & \metricwinner{0.791} & 4.568 \\
        \hline
        \rowcolor{metricrowa}
        \textbf{Average Run Time} & \metricwinner{3.429} & 16.471 \\
        \hline
        \rowcolor{metricrowb}
        \textbf{Average Wait Time} & \metricwinner{4.219} & 21.038 \\
        \hline
        \rowcolor{metricrowa}
        \textbf{Workload Changes} & \metricwinner{9} & 24 \\
        \hline
        \rowcolor{metricrowb}
        \textbf{TRF} & \metricwinner{2.332} & 1.000 \\
        \hline
        \rowcolor{metricrowa}
        \textbf{TiRF} & \metricwinner{1.805} & 1.000 \\
        \hline
        \rowcolor{metricrowb}
        \textbf{Average TiIF} & 0.919 & \metricwinner{1.000} \\
        \hline
        \rowcolor{metricrowa}
        \textbf{Average LPST} & -1.046 & \metricwinner{-0.674} \\
        \hline
    \end{tabular}
\end{table}

\begin{table}[!t]
    \centering
    \caption{\small InterQ comparison across two Large workloads for IBM LOCC, IonQ QComm, and Atom QComm. Green cells indicate the better value for that metric within each workload. Lower is better for $T_{\text{wait}}$, $T_{\text{run}}$, and $T_{\text{total}}$. For LPST, the value closer to $0$ is preferred.}
    \label{tab:interq_combined_two_workloads}
    \vspace{0.5ex}
    \setlength{\tabcolsep}{6.5pt}
    \renewcommand{\arraystretch}{1.18}
    \begin{tabular}{|c|c|c|c|c|}
        \hline
        \rowcolor{tableheaderdark}
        \textcolor{white}{\textbf{Mode}} &
        \textcolor{white}{\textbf{$T_{\text{wait}}$}} &
        \textcolor{white}{\textbf{$T_{\text{run}}$}} &
        \textcolor{white}{\textbf{$T_{\text{total}}$}} &
        \textcolor{white}{\textbf{LPST}} \\
        \hline

        \rowcolor{sectionrow}
        \multicolumn{5}{|c|}{\textbf{MQT 133 Jobs}} \\
        \hline
        \rowcolor{metricrowa}
        \cellcolor{tableheaderibm}\textcolor{white}{\textbf{IBM LOCC}} &
        \metricwinner{111.42} &
        \metricwinner{20.56} &
        \metricwinner{131.97} &
        -7.03 \\
        \hline
        \rowcolor{metricrowb}
        \cellcolor{tableheaderionq}\textcolor{white}{\textbf{IonQ QComm}} &
        901.74 &
        144.95 &
        1046.69 &
        \metricwinner{-1.56} \\
        \hline
        \rowcolor{metricrowa}
        \cellcolor{tableheaderatom}\textcolor{white}{\textbf{Atom QComm}} &
        663.17 &
        114.07 &
        777.25 &
        -1.67 \\
        \hline

        \rowcolor{sectionrow}
        \multicolumn{5}{|c|}{\textbf{Random 50 Jobs}} \\
        \hline
        \rowcolor{metricrowa}
        \cellcolor{tableheaderibm}\textcolor{white}{\textbf{IBM LOCC}} &
        \metricwinner{8.43} &
        \metricwinner{3.77} &
        \metricwinner{12.20} &
        \metricwinner{-0.27} \\
        \hline
        \rowcolor{metricrowb}
        \cellcolor{tableheaderionq}\textcolor{white}{\textbf{IonQ QComm}} &
        80.79 &
        67.10 &
        147.90 &
        -0.53 \\
        \hline
        \rowcolor{metricrowa}
        \cellcolor{tableheaderatom}\textcolor{white}{\textbf{Atom QComm}} &
        30.86 &
        22.64 &
        53.50 &
        -0.34 \\
        \hline
    \end{tabular}
\end{table}

\subsection{QComm Partitioning and Communication Cost}
\label{subsec:eval-qcomm}

In the quantum-link setting, aggregate schedule traces alone do not fully capture partitioning quality, since each partition also induces communication demand. We therefore report communication-characterization results that quantify remote-operation counts and teleportation-related qubit overhead.

We also compare IBM LOCC, IonQ QComm, and Atom QComm backends and evaluate modular execution in a cloud setting. Table ~\ref{tab:ibm-ionq-atom-11c-metrics} summarizes the 11-circuit workload used for this comparison, and the execution schedules for all 3 architectures are shown in \ref{fig:ibm-ionq-atom-11c-schedules}. The workload includes a 142-qubit circuit with the Job ID = 10 that cannot be placed monolithically on any of the evaluated backend configurations and is therefore partitioned in all three cases. In this experiment, the IonQ QComm configuration requires six QPUs to accommodate the partitioned workload, while the neutral-atom configuration achieves the best LPST, indicating the strongest fidelity estimate among the three architectures. The IBM LOCC method achieves the best runtime makespan as the QPUs are larger with 127 qubits, and classical reconstruction for these circuits were relatively negligible. We observe that the circuit cutting with LOCC can offer modular computing when the subcircuit sampling required is low, as in this case, as not many circuits were cut due to adaptive cutting technique based on the circuit requirements and the qubit utilization. 

\begin{figure}[!t]
    \centering
    \includegraphics[width=\linewidth]{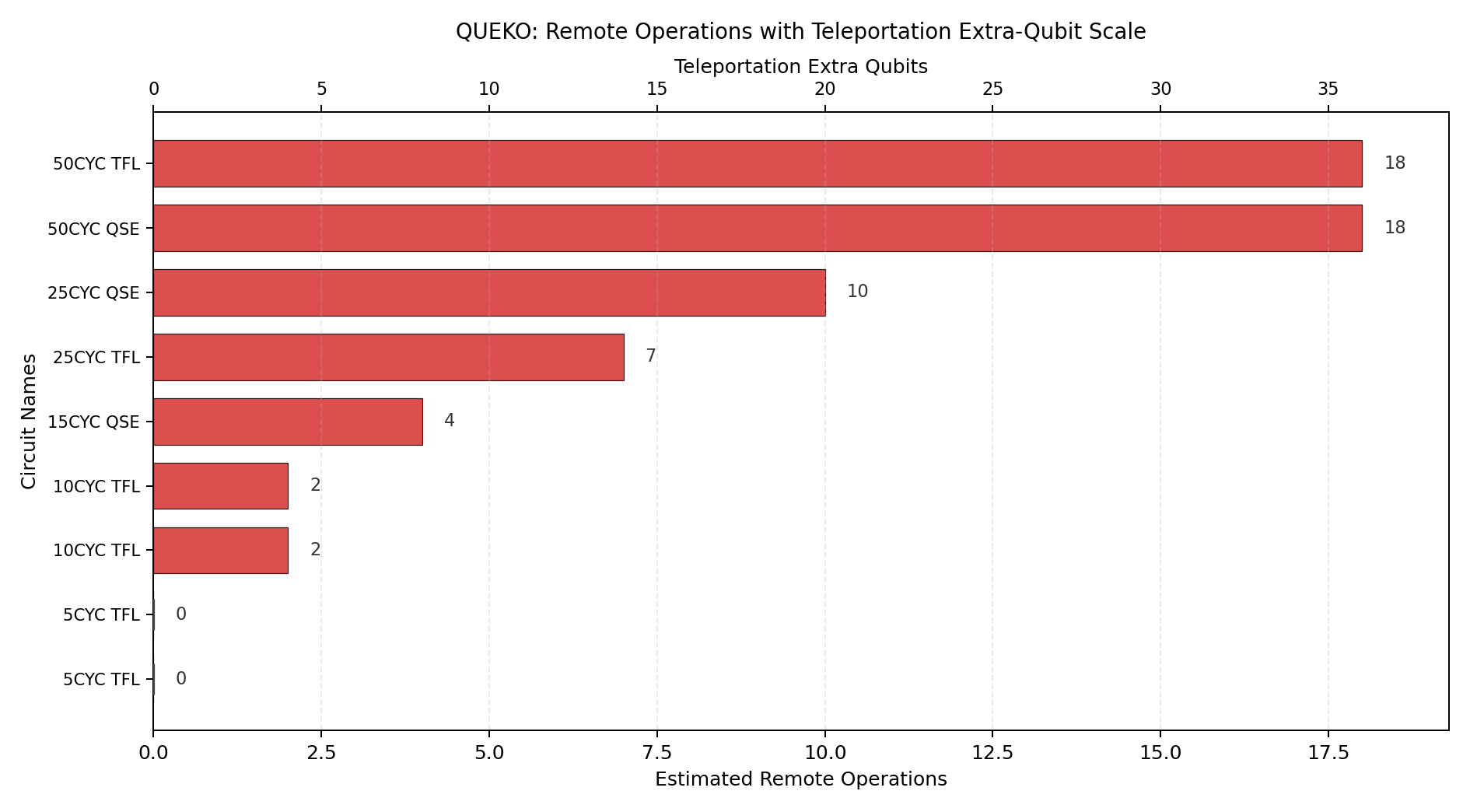}
    \caption{QUEKO remote operations versus extra qubits. The figure relates estimated remote-operation demand to teleportation-related qubit overhead for the QUEKO benchmark set.}
    \label{fig:interq_eval_queko_extra}
\end{figure}

\begin{figure*}[!t]
    \centering
    \subfloat[RevLib partitions and remote operations\label{fig:interq_eval_revlib_full}]{%
        \includegraphics[width=0.92\linewidth]{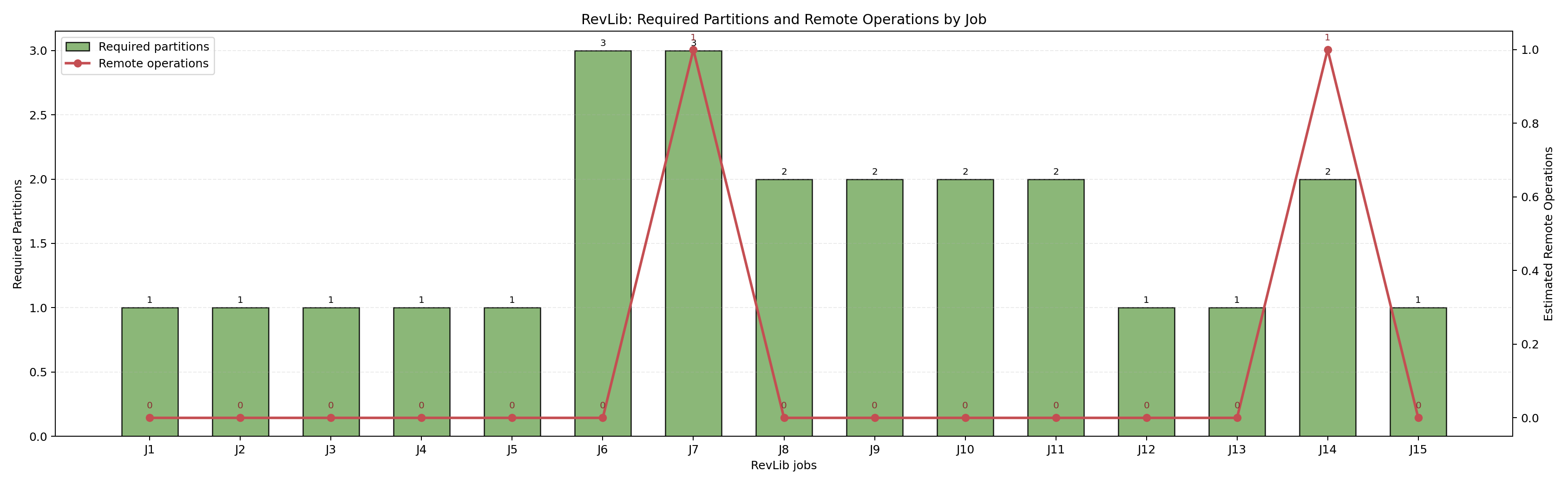}%
    }\\[0.8em]
    \subfloat[QUEKO partitions and remote operations\label{fig:interq_eval_queko_full}]{%
        \includegraphics[width=0.92\linewidth]{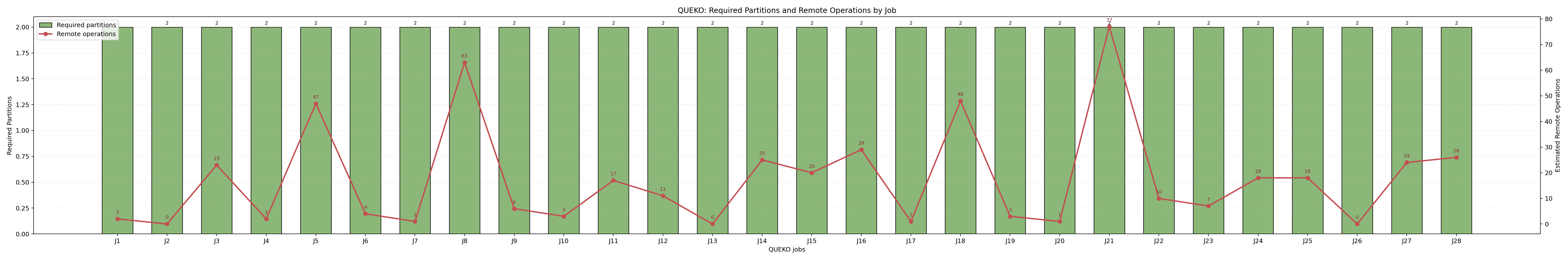}%
    }
    \caption{QComm partitioning and remote-operation summaries. The two plots show how required partition count and remote-operation burden vary across RevLib and QUEKO benchmark families.}
    \label{fig:interq_eval_qcomm_partitions}
\end{figure*}

\begin{figure}[!t]
    \centering
    \includegraphics[width=\linewidth]{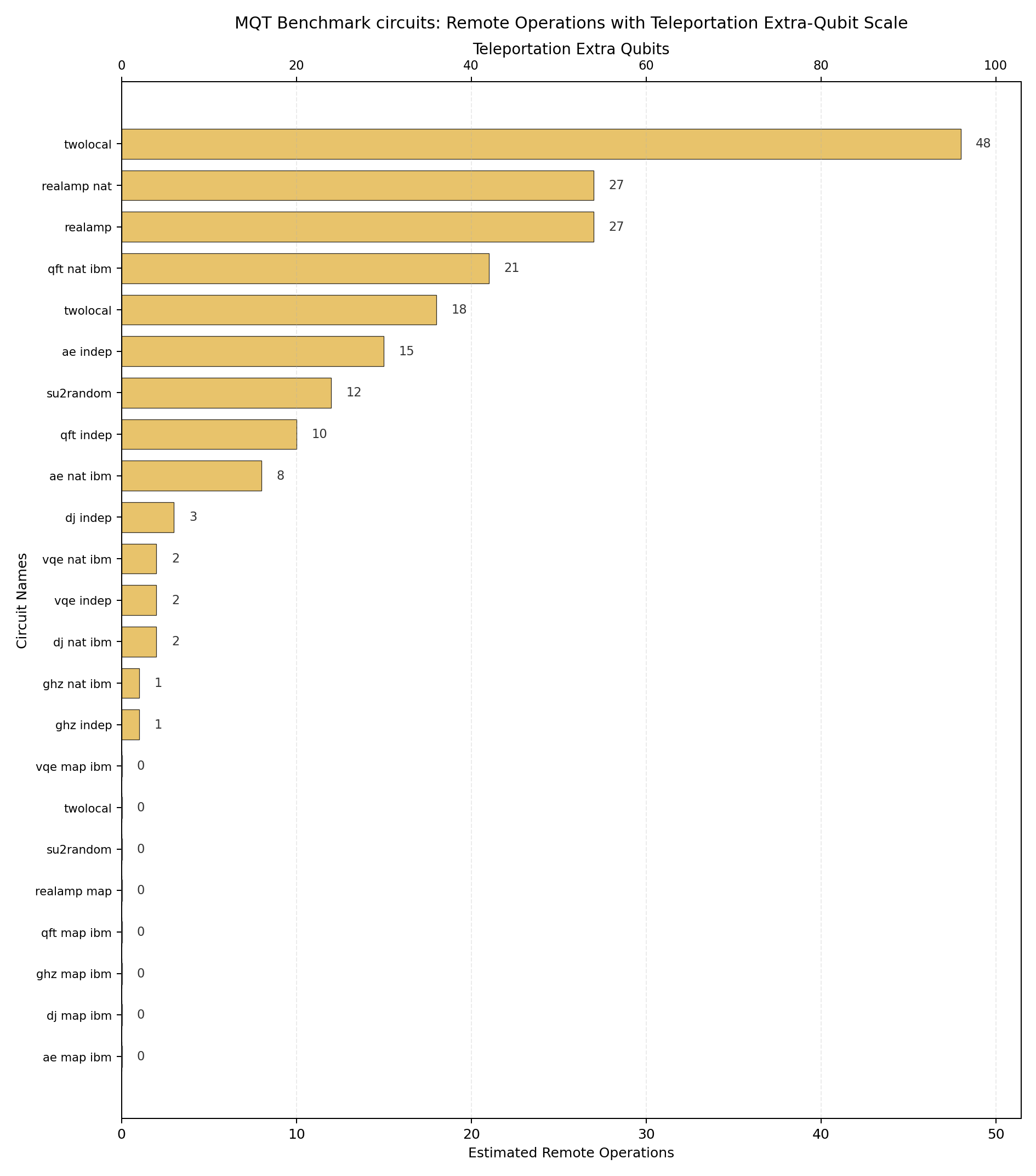}
    \caption{MQT small remote operations vs. extra qubits. The plot relates estimated remote operations to teleportation-related extra-qubit demand for the MQT small benchmark set.}
    \label{fig:interq_eval_mqt_extra}
\end{figure}

Figures~\ref{fig:interq_eval_qcomm_partitions}, \ref{fig:interq_eval_queko_extra}, and~\ref{fig:interq_eval_mqt_extra} provide complementary views of QComm-induced communication cost. Figure~\ref{fig:interq_eval_qcomm_partitions} reports benchmark-level partition and remote-operation summaries for RevLib and QUEKO. Figures~\ref{fig:interq_eval_queko_extra} and~\ref{fig:interq_eval_mqt_extra} show the relationship between estimated remote operations and teleportation-related qubit overhead for the QUEKO and MQT small benchmark sets, respectively.

Table~\ref{tab:interq_combined_two_workloads} shows the effect of large job queues. For the 133-job MQT workload, IBM LOCC achieves the lowest average runtime by exploiting parallel execution across mixed-width circuits, with only a modest LPST reduction. IonQ-style devices, with smaller simulated capacity, support fewer concurrent jobs and therefore have higher runtime but the best LPST.

For the 50-job random-circuit workload, IonQ and neutral-atom configurations require more partitioning, increasing distributed-execution overhead. Consequently, IBM LOCC achieves the best average makespan and LPST, as highlighted in green.

\section{Conclusion}
\label{sec:conclusion}

This paper presented InterQ, a communication-aware scheduling framework for modular quantum execution. InterQ jointly considers job placement, circuit partitioning, parallel grouping, and communication-resource constraints across LO, LOCC, and quantum-link-based QComm execution. Rather than treating communication as a secondary placement cost, InterQ incorporates synchronization delay, cut-induced overhead, remote-operation demand, Bell-pair availability, and link contention directly into scheduling decisions.

The evaluation shows that interconnect assumptions strongly affect scheduling behavior and system-level trade-offs. Classical-link LOCC execution improves feasibility by decomposing oversized circuits into causally ordered fragments and exposing parallelism when backend capacity is underutilized. Quantum-link execution is instead shaped by remote-operation demand and shared-link availability, making link-aware placement essential for maintaining throughput. Across workload settings, neutral-atom modular QPUs provide the strongest fidelity profile, superconducting systems achieve the lowest runtimes, and trapped-ion systems offer a balanced intermediate trade-off between fidelity and makespan.

Overall, InterQ shows that scalable modular quantum computing requires scheduling policies aware of both computation and communication. As modular platforms mature, InterQ provides a foundation for studying distributed-resource scalability, including communication links, Bell-pair generation capacity, qubit availability, and heterogeneous QPU characteristics in growing quantum cloud systems.

\bibliographystyle{IEEEtran}
\bibliography{ref}

\end{document}